\documentclass[%
 pra,twocolumn,
superscriptaddress,
 amsmath,amssymb,
 aps,
]{revtex4-1}

\usepackage{braket} 

\usepackage{float} 

\usepackage{graphicx}
\usepackage{dcolumn}
\usepackage{bm}
\usepackage{subfigure}
\usepackage{breqn}

\graphicspath{{figs/}}

\makeatletter
\let\cat@comma@active\@empty
\makeatother

\begin{document}

\preprint{APS/123-QED}

\title{Tuning terahertz transitions in a double-gated quantum ring}

\author{T. P. Collier}
\email{tpc207@exeter.ac.uk}
\affiliation{School of Physics, University of Exeter, Stocker Road, Exeter EX4 4QL, United Kingdom}

\author{V. A. Saroka}
\email{v.saroka@exeter.ac.uk}
\affiliation{School of Physics,  University of Exeter, Stocker Road, Exeter EX4 4QL, United Kingdom}
\affiliation{Institute for Nuclear Problems, Belarusian State University, Bobruiskaya 11, 220030 Minsk, Belarus}

\author{M. E. Portnoi}
\email{m.e.portnoi@exeter.ac.uk}
\affiliation{School of Physics, University of Exeter, Stocker Road, Exeter EX4 4QL, United Kingdom}

\date{16 October 2017}

\begin{abstract}

We theoretically investigate the optical functionality of a semiconducting quantum ring manipulated by two electrostatic lateral gates used to induce a double quantum well along the ring. The well parameters and corresponding inter-level spacings, which lie in the THz range, are highly sensitive to the gate voltages. Our analysis shows that selection rules for inter-level dipole transitions, caused by linearly polarized excitations, depend on the polarization angle with respect to the gates. In striking difference from the conventional symmetric double well potential, the ring geometry permits polarization-dependent transitions between the ground and second excited states, allowing the use of this structure in a three-level lasing scheme.


\end{abstract}

\pacs{Valid PACS appear here}
\maketitle

\section{\label{sec:Intro}Introduction}

The THz regime is a narrow region of the electromagnetic spectrum for which practical technologies lack the ability to produce or detect coherent radiation. Aptly named the THz gap, bridging it constitutes one of the most formidable problems of modern applied physics. THz devices hold great potential in applications across diverse fields of science - examples include non-invasive biomedical imaging, submillimeter astronomy, and stand-off detection of plastic explosives \cite{Williams2006, Lewis2014, Siegel2002}. As such, interest in proposing efficient and portable THz devices has rapidly grown over the past couple of decades. 

The literature provides a variety of proposals for practical THz devices. Amongst the most heavily researched aspirants are semiconductor nanostructures hosting multiple quantum wells, such as double quantum wells \cite{Roskos1992,
Leo1991}, and quantum cascade lasers \cite{Faist1994,Williams2007,Fathololoumi2012}. The above mentioned double quantum well (DQW) structures are realized in heterostructures of alternating semiconductor layers with different band gaps. The shapes of these double well potentials are therefore intrinsic to the specific heterostructure and, whilst robust, offer limited ability to manipulate the shapes of the potentials without the use of large external fields. Promising candidates also exist in the form of non-simply-connected nanostructures, such as carbon nanotubes \cite{Portnoi2009,Batrakov2009,Portnoi2015,Hartmann2014}, double-layer graphene \cite{Batrakov2012} or quantum rings \cite{Alexeev2012a,Alexeev2012b,Alexeev2012c}. The appeal in using these latter structures for THz devices, in lieu of those former, lies in their greater tunability with external fields. 

A quantum ring (QR) represents the simplest example available for study of such a non-simply-connected system. These novel nanostructures have already been shown to exhibit a range of fascinating phenomena - predominantly resulting from the Aharonov-Bohm effect \cite{Alexeev2012a,Alexeev2012b,Alexeev2012c, Lorke2000, Chaplik1995, Romer2000, Llorens2001,Govorov2002,Maslov2003, Lee2004, daSilva2005, Planelles2006, Fischer2009, Teodoro2010,González-Santander2011,Roy2012,Climente2014,Simonin2014,Koshelev2015,Radu2015, PhysicsOfQRBook}. Exploiting Aharonov-Bohm-related properties requires trapping magnetic flux in the ring annulus, with flux equal to an odd number of one-half of the flux quantum enhancing the corresponding effects. A drawback of utilizing the Aharonov-Bohm effect in such rings for THz applications is therefore the amount of magnetic flux required. A quantum ring with a typical energy scale within the THz regime requires a radius on the order of 10nm. While constructing a nanostructure with these dimensions is certainly experimentally feasible \cite{Lorke2000,Lee2004}, to achieve the necessary flux piercing the ring would require a magnetic field $B\approx 10$T - limiting our ability to propose compact and portable QR-based THz devices. The aforementioned proposals \cite{Alexeev2012a,Alexeev2012b,Alexeev2012c} were based on lifting the degeneracy between QR energy levels occurring at half-integer flux quanta by means of an external electric field. Here we propose an alternative set-up with tunable energy levels without resorting to excessively large magnetic fields.

In this paper we analyze a quantum ring system with two electrostatic lateral gates and exploit its double-connectedness in the absence of magnetic fields. In what follows we theoretically investigate the influence of these lateral gates and show that this system is analogous to a DQW. Consequently, the control of energy separations can be achieved via tuning the double well parameters with experimentally attainable gate voltages. We highlight a  significant difference between this double-gated ring system and the conventional DQW, namely that intra-band optical transitions between the ground and second excited states are dependent on the polarization angles of incident linearly polarized light.

\section{\label{sec:TwoChargedWires}Theoretical Model}

\begin{figure}[h!]
\begin{center}
\includegraphics[width= 0.48\textwidth]{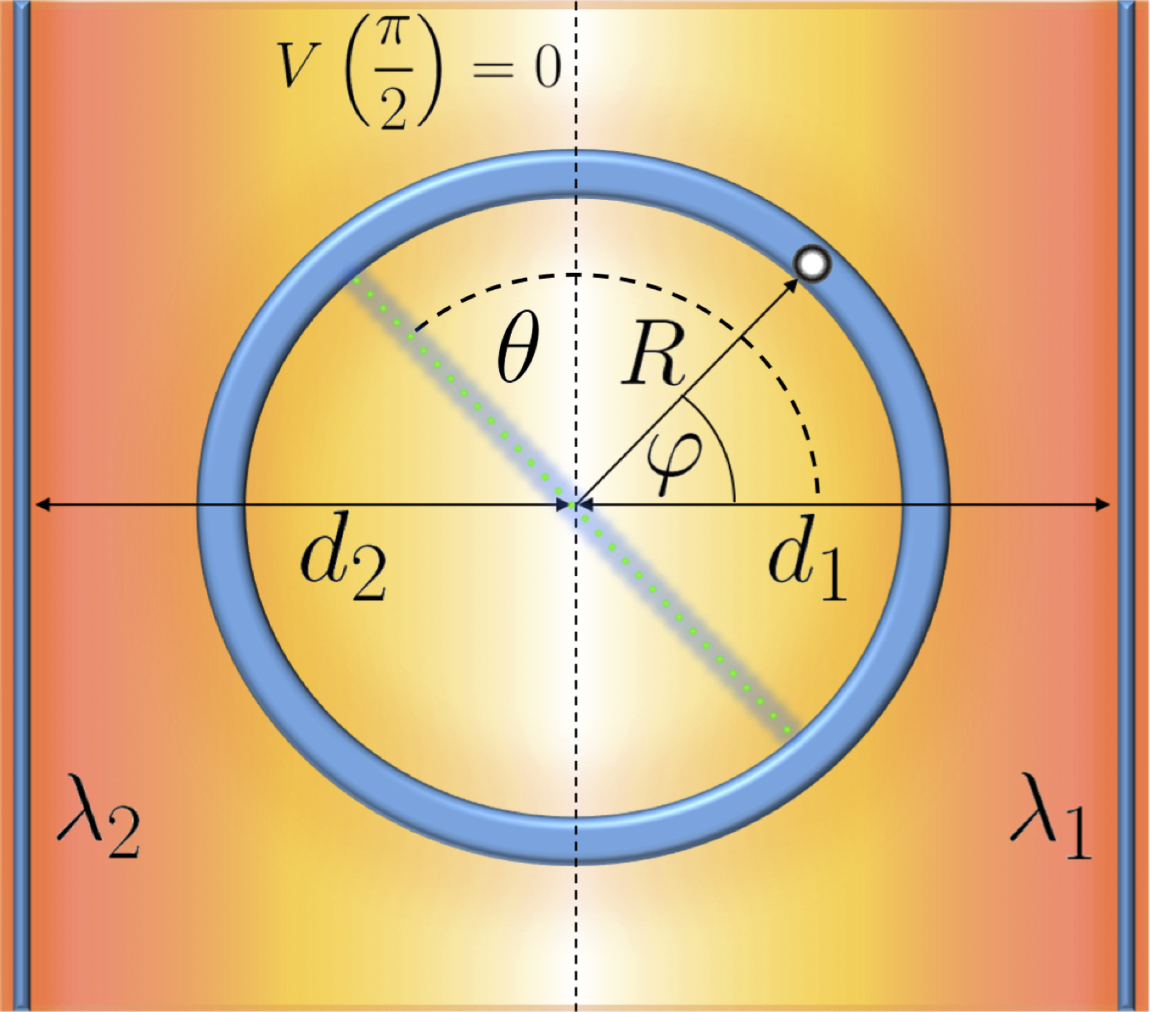}
\end{center}
\caption{\label{fig:MainParameters} (Colour online) A diagram of the system's geometry and parameters. $R$ is the ring radius, $d_1$ and $d_2$ are the distances of the charged wires from the ring center, with charge densities $\lambda_1$ and $\lambda_2$ respectively. The angular position along the ring is given by $\varphi$, measured from the horizontal axis as shown. The dotted green line is the projection of the linearly polarized radiation onto the plane of the ring, where $\theta$ is the angle between this projection and the horizontal axis.
}
\end{figure}

We consider the system of a semiconductor QR hosting a single electron between two in-plane gates. We have considered two models for theses gates; point charges and infinitely long charged wires, with both yielding qualitatively similar conclusions. In this paper we present results for the gates in the form of the latter because this set-up can accommodate linear arrays of QRs between the gates. The geometry of the system is shown in Fig.~\ref{fig:MainParameters}. We treat the QR as infinitely narrow, which is a reasonable approximation for a ring with mean radius considerably larger than its width and height \cite{González-Santander2011}. The Hamiltonian within the effective mass approximation reads, 
\begin{equation}
\mathcal{H} =  -\frac{\partial^2}{\partial\varphi^2} +V(\varphi),
 \label{eq:Hamiltonian}
\end{equation}
 where $V(\varphi)$ is the external potential. The angular position along the ring $\varphi$ is taken from the horizontal axis as shown in Fig.~\ref{fig:MainParameters}. The electrostatic potential along the ring due to an individual charged wire is
\begin{equation}
 \Phi_i(\varphi) = - \lambda_i k \ln\left[\frac{d_i+(-1)^{i}R\cos(\varphi)}{d_i}\right],
 \label{eq:potential_line}
\end{equation}
where $i=1,2$ labels the wires, $\lambda_i$ is the magnitude of the linear charge density on a wire, $d_i$ is its corresponding distance from the centre of the ring, and $k = 1/2\pi \epsilon$ with $\epsilon$ the absolute permittivity. We have defined zero potential to be along the parallel line passing through the centre of the ring, $V\left(\frac{\pi}{2} \right)=0$, as shown. The dimensionless potential felt by an electron of charge $-e$ due to the presence of negatively charged wires is given by $V(\varphi) = e(\Phi_1 + \Phi_2)/\varepsilon_1(0)$, where $\varepsilon_1(0) = \hbar^2/2\mu R^2$ is the energy scale for a ring with radius $R$ and electron effective mass $\mu$. The $\varepsilon_1(0)$ notation is similar to that introduced in Ref.~\cite{Alexeev2012a}. For a typical semiconducting ring with $R=20$nm and $\mu=0.05 m_e$ this gives an energy scale $\varepsilon_1(0) \simeq 2$meV  which corresponds to 0.5THz. Under an appropriate expansion whereby the radius is much smaller than the distances from the ring center to the gates, $R/d_i \ll 1$, the potential may be expressed as
\begin{equation}
 V(\varphi) = 2\beta \frac{d}{R}(1 - \gamma)\cos(\varphi) + \beta \gamma \cos(2\varphi),
 \label{eq:potential}
\end{equation}
where $\beta = \lambda ke R^2 / 2d^2 \varepsilon_1(0)$ characterizes the strength of the potential. We have defined the asymmetry term $\gamma$ as the ratio of the distances divided by the ratio of the charge densities  $\gamma = d_1 \lambda_2 / d_2 \lambda_1$ and dropped the subscripts for convenience, $\lambda_1 \equiv \lambda$ and $d_1 \equiv d$. It should be noted that the inevitable asymmetry caused by the difficulty in maintaining $d_1 = d_2$ can be compensated by manipulating $\lambda_1$ and $\lambda_2$. The first term in Eq.~\eqref{eq:potential} is dominant unless $|1-\gamma|\, \sim R/d$, otherwise the potential effectively reduces to the case of a ring under the influence of a lateral electric field  \cite{Llorens2001,Radu2015, Alexeev2012a, Alexeev2012b, Alexeev2012c}.

The $2\pi$-periodic wavefunctions of the Hamiltonian in zero potential are given by $\Psi_m(\varphi)=\exp(im\varphi)/\sqrt{2\pi}$, where $m$ is the angular momentum quantum number. Upon introduction of the potential \eqref{eq:potential} the axial symmetry is broken and electron states with different angular momentum are mixed, consequently $m$ is defunct as a quantum number. Ergo, we seek solutions to Eq.~\eqref{eq:Hamiltonian} as a linear combination of these basis functions
\begin{equation}
   \Psi_n(\varphi) =\sum_{m=-\infty}^{m=+\infty} c_m^{(n)}e^{im\varphi}, 
    \label{eq:basisfunctions}
\end{equation}
which maintains the required $2\pi$-periodicity in $\varphi$, and $n$ labels the $n^{\text{th}}$ eigenstate. We can make use of the orthogonality of exponential functions by multiplying the resulting expression by $e^{im'\varphi}/2\pi$ and integrating with respect to $\varphi$, where $m'$ is some integer. The corresponding infinite set of simultaneous equations for the coefficients $c_m^{(n)}$ reads as
\begin{multline}
  (m^2-a_n)c_m^{(n)}+\beta \frac{d}{R}(1- \gamma)\left(c_{m-1}^{(n)}+c_{m+1}^{(n)} \right)\\ + \frac{\beta \gamma}{2}\left(c_{m-2}^{(n)}+c_{m+2}^{(n)} \right)=0,
    \label{eq:simultaneousequations}
\end{multline}
and represents an infinite penta-diagonal matrix. Here $a_n = \varepsilon_n / \varepsilon_1(0)$ is the dimensionless $n^{\text{th}}$ eigenenergy. We see from Eq.~\eqref{eq:simultaneousequations}, that when $\gamma=1$ only the states with $\Delta m=\pm2$ are mixed, whereas the uniaxial direction established by $\gamma\ne1$ also mixes states with $\Delta m=\pm1$. Truncating and numerically diagonalising the matrix \eqref{eq:simultaneousequations} yields the eigenenergies and their corresponding coefficients $c_m^{(n)}$, from which the wavefunctions are given via Eq.~\eqref{eq:basisfunctions}. For the remainder of this paper we apply a truncation at $|m|=13$, any increase in matrix size yields no noticeable change in the lowest eigenenergies. 
\begin{figure*}[t]
\begin{center}
\includegraphics[width=\textwidth]{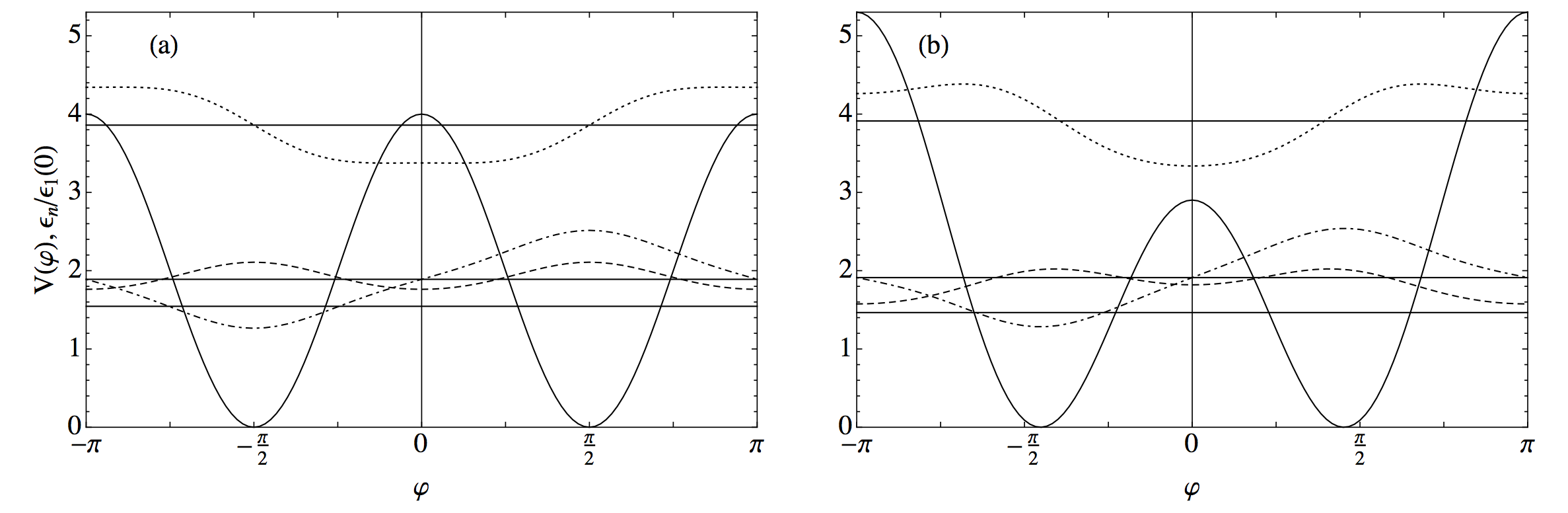}
\end{center}
\caption{\label{fig:LifshitzArse} Potential and lowest three energy levels  with the corresponding wave functions sketched on top for $\beta =2 $ and $d/R = 100$: (a) is for $\gamma = 1$, and (b) is for $\gamma=1.003$. The potentials and energy levels have been shifted by a constant to bring the well minima to zero.}
\end{figure*}

\section{\label{sec:DoubleWell}Tuneable Double Quantum Well}

\begin{figure}[t]
\begin{center}
\includegraphics[width= 0.49\textwidth]{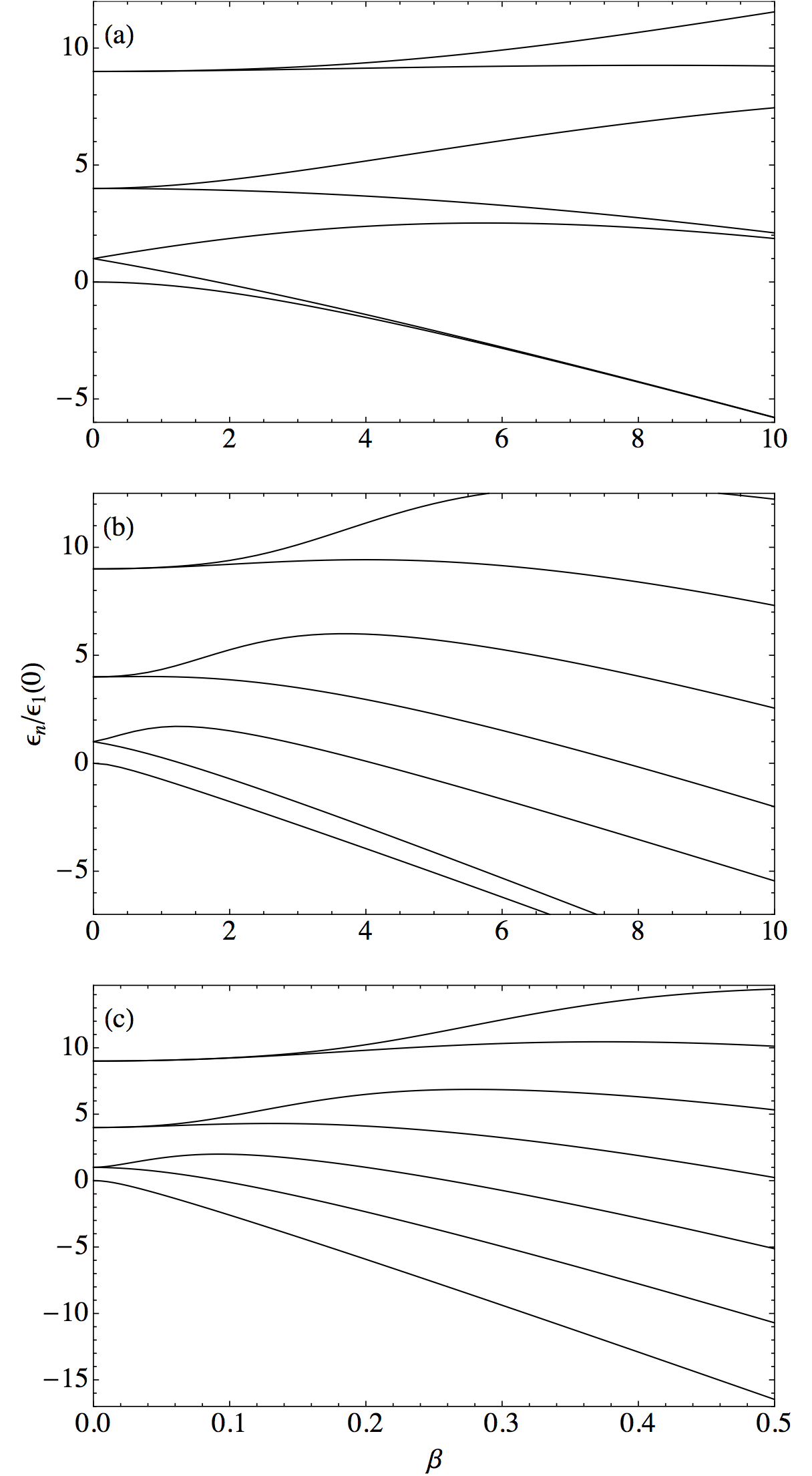}
\end{center}
\caption{\label{fig:energydispersions} Lowest lying eigenenergies plotted as a function of $\beta$ for reasonable values of asymmetry parameter: (a) is for $\gamma=1$, and (b) is for  $\gamma=1.009$, and a large value (c) is for $\gamma=1.2$.}
\end{figure}

The formation of a double quantum well (DQW) has only been discussed in QRs subject to exceedingly large in-plane magnetic fields \cite{Planelles2006}, however here we show that DQW solutions occur for a QR subject to the double gate potential \eqref{eq:potential} with easily attainable voltages. Fig.~\ref{fig:LifshitzArse} illustrates the lowest three levels, with the corresponding wave functions superposed on top, for $\beta=2$ and $d/R = 100$ (Fig.~\ref{fig:LifshitzArse}(a) is $\gamma=1$ and Fig.~\ref{fig:LifshitzArse}(b) is $\gamma = 1.003$). The potentials and energy levels have all been shifted by a constant so that the well minima are at zero. The potential strength $\beta$ and asymmetry parameter $\gamma$ permit control over the form of the DQW, thereby allowing control of the energy levels solely via the voltages applied to the gates (assuming the spatial parameters are kept fixed). Indeed, the positions of the well minima, $\pm\varphi_0$, are governed by $\gamma$ such that $\varphi_0 = \textrm{Arccos}\left[\frac{d}{2 R}(\gamma -1)/\gamma\right]$. The well maxima occur at $\varphi=0,\pm\pi$ and are given via Eq.~\eqref{eq:potential}. As the potential strength $\beta$ increases, successive energy states pair up and asymptotically approach a doubly degenerate state, with the inter-level splitting decreasing for higher $\gamma$. This is reflected in the upper two panels of Fig.~\ref{fig:energydispersions}, which shows the lowest lying energy levels as a function of $\beta$ for several values of $\gamma$. When $\gamma=1.009$, the DQW solutions form, albeit at a higher value of $\beta$ in comparison with when $\gamma=1$. Hence, the asymmetry parameter $\gamma$ acts to reduce the sensitivity of the system. The bottom panel of Fig.~\ref{fig:energydispersions} depicts the dispersion for $\gamma=1.2$ and resembles that of a QR in a lateral electric field \cite{Alexeev2012a} wherein DQW solutions are not formed. Thus, we demonstrate that the lateral field setting can be treated on the same footing as a double-quantum-well-on-a-ring problem.

\begin{figure}[H]
\begin{center}
\includegraphics[width= 0.48\textwidth]{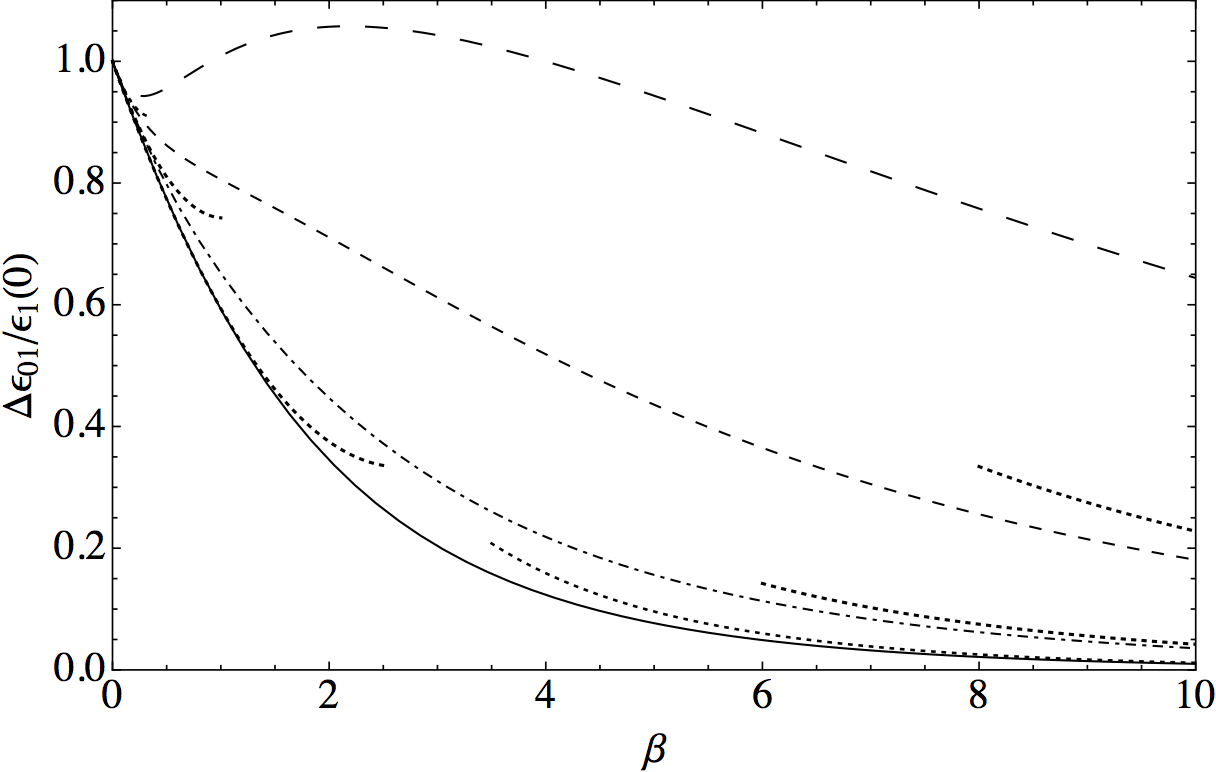}
\end{center}
\caption{\label{fig:energySeparations} Energy separation between ground state and first excited state as a function of $\beta$. The solid curve is for $\gamma=1$, subsequently the dot-dashed curve is for $\gamma=1.003$, the small-dashed curve is for $\gamma=1.006$, and $\gamma=1.009$ is the large-dashed curve. Dotted lines represent approximate results obtained by perturbation theory and the WKB method, not visible for $\gamma=1.009$ within the plotted range and scale of $\beta$.}
\end{figure}

For small potential strengths, $(\beta \ll 1)$ and reasonable asymmetry parameters $(|1-\gamma|\, \sim R/d)$ we can express the energy separation between the ground state and the first excited state, up to second order in $\beta$, via Rayleigh-Schr\"{o}dinger perturbation theory as
\begin{equation}
  \Delta\varepsilon_{01} = \varepsilon_1(0)\left(1- \frac{\beta \gamma}{2} + \frac{3\beta^2\gamma^2}{32} + \frac{5d^2}{3R^2}\beta^2(1- \gamma)^2\right).
    \label{eq:SmallStrengthSplitting}
\end{equation}
As we increase $\beta$, higher order terms dominate and we resort to a non-perturbative approach. In the limit of high $\beta$, the lowest two states are well described by odd and even combinations of harmonic oscillator ground states localized in the individual wells. Tunneling between these degenerate states through the two potential barriers around the ring results in an energy splitting. To describe this tunnelling-dependent splitting we use the WKB approximation, adapted to a DQW on a ring, with the appropriate pre-factor attributed to the ground state splitting \cite{Garg2000,LandLpr3}. Alternative tunneling-dependent methods would be also appropriate \cite{Garg2000,MullerKirsten2006}. The separation between the ground and the first excited state, adapted to angular coordinates is given as
\begin{equation}
  \Delta\varepsilon_{01}^{\textsc{\tiny{(WKB)}}} = \frac{\hbar \omega}{\sqrt{e\pi}}\left( e^{-\Phi(\varphi_c)}+e^{-\Phi(\varphi^{\prime}_c)} \right),
    \label{eq:WKBSplitting}
\end{equation}
where $\omega$ is the oscillator frequency of an individual well. The classical turning points $\varphi_c$ and $\varphi^{\prime}_c$ are the positive solutions of $V(\varphi)=E$ where $E$ is the harmonic oscillator ground state energy in an individual well. The two terms on the right-hand side of Eq.~\eqref{eq:WKBSplitting} represent separate contributions from tunneling through the barriers centred on $\varphi=0$ and $\varphi=\pi$, hence the exponents are given via 
\begin{equation}
 \Phi(\varphi_c) = \int_{-\varphi_c}^{\varphi_c} p(\varphi) \; d\varphi, \qquad \quad \Phi(\varphi^{\prime}_c) = \int_{\varphi^{\prime}_c}^{2\pi-\varphi^{\prime}_c} p(\varphi) \; d\varphi,
 \label{eqn:momentum}
\end{equation}
where $p(\varphi)=\sqrt{V(\varphi)-E}$. Results for the energy separation between the ground state and first excited state as a function of $\beta$ are plotted in Fig.~\ref{fig:energySeparations} for several values of $\gamma$. Results from Eq.~\eqref{eq:SmallStrengthSplitting} and Eq.~\eqref{eq:WKBSplitting} are also plotted as dotted lines for $\gamma=1$, $1.003$, and $1.006$. When $\gamma=1$, the approximate perturbation theory result agrees well with the exact calculations for potential strengths as high as $\beta \lesssim 1$. The WKB result then agrees from $\beta \gtrsim 6$. As $\gamma$ increases, we see that the approximate perturbation theory expressions agree only for lower $\beta$, likewise the WKB results agree for higher $\beta$. This is because the $\cos(\varphi)$ term in the potential \eqref{eq:potential} acts to repulse the first excited state from the ground state, as can be seen from the last term in Eq.~\eqref{eq:SmallStrengthSplitting}. Indeed, for large enough values of $\gamma$, after an initial decrease, the separation then increases with increasing $\beta$ before finally decreasing once more (as can be seen by $\gamma=1.009$ in Fig.~\ref{fig:energySeparations}). When only the ground state is confined, there is an initial decrease in energy separation with respect to $\beta$. Then, upon confining the first excited state by the larger potential barriers, the $\cos(\varphi)$ term acts as a repulsion between the bottom two energy levels. With increasing $\beta$, both states drop deeper into the potential and are confined by the smaller potential barrier. Correspondingly, $\Delta\varepsilon_{0,1}$ decreases as the states approach an asymptotically degenerate state.

\section{\label{sec:EnergySplitting}Dipole Transition Selection Rules}

Here we study the radiative intra-band transitions in our double-gated QR. Let us consider linearly polarized radiation incident onto the QR, with $\boldsymbol{ \eta }$ representing the projection of the polarization vector onto the plane of the ring, and restrict ourselves solely to dipole optical transitions. The transition dipole matrix element $P_{if}(\theta)=\braket{ f|\boldsymbol{ \eta } \mathbf{\hat{P}}|i }$ dictates the transition rate between initial ($i$) and final ($f$) single-electron states $\Gamma_{if}\propto|P_{if}(\theta)|^2$, where we have defined the dipole moment operator as $\mathbf{\hat{P}}$. For our model of an infinitely narrow ring, this matrix element takes the form \cite{Alexeev2012a}
\begin{equation}
  P_{if}(\theta) = eR\int \Psi_f^{\ast}\Psi_i\cos(\theta - \varphi)d\varphi,
    \label{eq:TransitionDipoleMatrixElement}
\end{equation}
where $e$ is the elementary charge and $\theta$ is the angle between the projection of linearly-polarized radiation onto the plane of the QR and the horizontal axis, as shown in Fig.~\ref{fig:MainParameters}. 

As the potential induced by the side gates is symmetric about $\varphi=0$, the wave functions given by Eq.~\eqref{eq:basisfunctions} must be either odd or even about $\varphi=0$, and it is convenient to separate them by parity. We begin by considering the system with $\gamma=1$, corresponding to mixing states with angular momentum quantum numbers differing by $\Delta m = \pm2$, as can be seen from Eq.~\eqref{eq:simultaneousequations}. The additional symmetry about $\varphi=\pi /2$ permits both $\pi$-periodic and $2\pi$-periodic solutions, which allows us to express the lowest 4 electronic states as
\begin{subequations}
\begin{equation}
  \Psi_0(\varphi) = \frac{1}{\sqrt{\pi}}\sum_{j=0}A^{(0)}_{2j}\cos(2j\varphi),
    \label{eq:GSsymmetricSmallB}
\end{equation}
\begin{equation}
  \Psi_1(\varphi) = \frac{1}{\sqrt{\pi}}\sum_{j=0}B^{(1)}_{2j+1}\sin[(2j+1)\varphi],
    \label{eq:1SsymmetricSmallB}
\end{equation}
\begin{equation}
  \Psi_2(\varphi) = \frac{1}{\sqrt{\pi}}\sum_{j=0}A^{(2)}_{2j+1}\cos[(2j+1)\varphi],
    \label{eq:2SsymmetricSmallB}
\end{equation}
\begin{equation}
  \Psi_3(\varphi) = \frac{1}{\sqrt{\pi}}\sum_{j=0}B^{(3)}_{2j+2}\sin[(2j+2)\varphi],
    \label{eq:3SsymmetricSmallB}
\end{equation}
\end{subequations}
where we have included the $1/\sqrt{\pi}$ pre-factors for later convenience. Incidentally, as the stationary Schr\"{o}dinger equation is the Mathieu equation \cite{Mathieu1868}, these are periodic Mathieu functions \cite{AbramowitzAndStegun72} (although we have omitted the conventional $1/\sqrt{2}$ pre-factor for the ground state coefficients). These fourier coefficients can be found numerically via diagonalization of Eq.~\eqref{eq:simultaneousequations} or found via approximate methods for small $\beta$ \cite{AbramowitzAndStegun72}. Substituting Eq.~\eqref{eq:GSsymmetricSmallB}, \eqref{eq:1SsymmetricSmallB}, and \eqref{eq:2SsymmetricSmallB} into Eq.~\eqref{eq:TransitionDipoleMatrixElement} as the respective electron wave functions $\Psi_{i}$ and $\Psi_{f}$, we find for transitions between the three lowest states
\begin{subequations}
\begin{equation}
  P_{01}(\theta) = eR\sin(\theta)\sum_{j,k=0}\frac{B^{(1)}_{2j+1}A^{(0)}_{2k}}{2-\delta_{k,0}}\left(\delta_{j,k} - \delta_{j,k -1} \right),
    \label{eq:TransitionDipoleMatrixElementSymmetric01}
\end{equation}
\begin{equation}
  P_{02}(\theta) = eR\cos(\theta)\sum_{j,k=0}\frac{A^{(2)}_{2j+1}A^{(0)}_{2k}}{2-\delta_{k,0}}\left(\delta_{j,k} + \delta_{j,k -1} \right),
    \label{eq:TransitionDipoleMatrixElementSymmetric02}
\end{equation}
\begin{equation}
P_{12}(\theta)=0.
\label{eq:TransitionDipoleMatrixElementSymmetric12}
\end{equation}
\end{subequations}
Remarkably, optical transitions are allowed between the ground state and both the first and second excited state, while the transition between these excited states is forbidden for $\gamma=1$. This represents a clear difference from typical DQW heterostructures, where optical transitions are only allowed between states of opposite parity, such as between the second and first excited state, and transitions between states of the same parity are typically forbidden. Here however, due to the ring geometry, dipole transitions between the ground and the second excited state are allowed, with $\Gamma_{02}$ reaching its maximum value when induced by radiation polarized parallel to horizontal axis $\theta = 0$. This is also different from what is expected from the periodic and hard wall boundary conditions in carbon nanotubes and graphene nanoribbons~\cite{Saroka2017}. We can understand this new feature by considering the unperturbed QR system, which has a zero ground state energy with constant wavefunction and a doubly degenerate first excited state (reflecting rotational symmetry about the ring). Optical selection rules allow transitions from the ground state to one of these adjacent states only, corresponding to states with angular momenta quantum number differing by unity. Any breaking of axial symmetry reveals the zero-th order wavefunctions for these double degenerate states to be $\cos(\varphi)$ and $\sin(\varphi)$. Thus, perturbing the QR with the double well potential we see that the state corresponding to $\sin(\varphi)$, with its nodes centred on the potential barriers, is less influenced by the perturbation and remains as the first excited state, whereas the state corresponding to $\cos(\varphi)$, having extrema at the barriers,  is repulsed by the potential barriers and becomes the second excited state. This is evident by the first terms of the series from Eq.~\eqref{eq:1SsymmetricSmallB}, and ~\eqref{eq:2SsymmetricSmallB}. Allowed optical transitions therefore persist between the ground state and the next two excited states for the double-gated QR, with the polarization angle-dependent selection rules manifesting from the ring geometry and the defined parity from the perturbation.
\begin{figure}[t]
\begin{center}
\includegraphics[width= 0.48\textwidth]{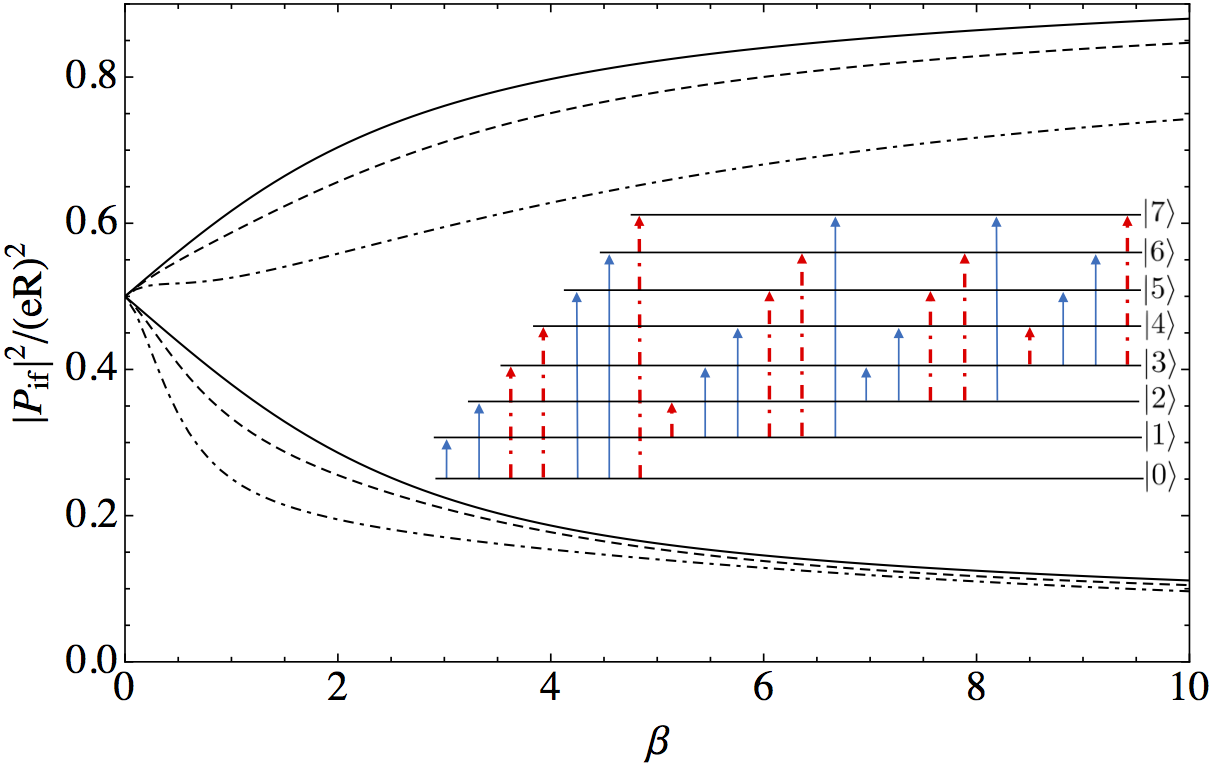}
\end{center}
\caption{\label{fig:dipolematrixelements0102} (Colour online) Square of dimensionless transition dipole matrix element between initial ($i$) and final ($f$) single-electron states plotted as a function of $\beta$ for different values of $\gamma$. The upper branch is $|P_{01}|^2/(eR)^2$ at $\theta=\pi/2$ and the lower branch is $|P_{02}|^2/(eR)^2$ at $\theta=0$. The full lines denote the results for $\gamma=1$, dashed lines are for $\gamma=1.003$, and dot-dashed lines are for $\gamma=1.006$. The inset is a schematic showing the optical selection rules between the eight lowest energy eigenstates. The full blue arrows are the allowed transitions for when $\gamma=1$, the red dot-dashed arrows are forbidden transitions which become allowed when $\gamma \ne 1$.}
\end{figure}

\begin{figure}[h]
\begin{center}
\includegraphics[width= 0.48\textwidth]{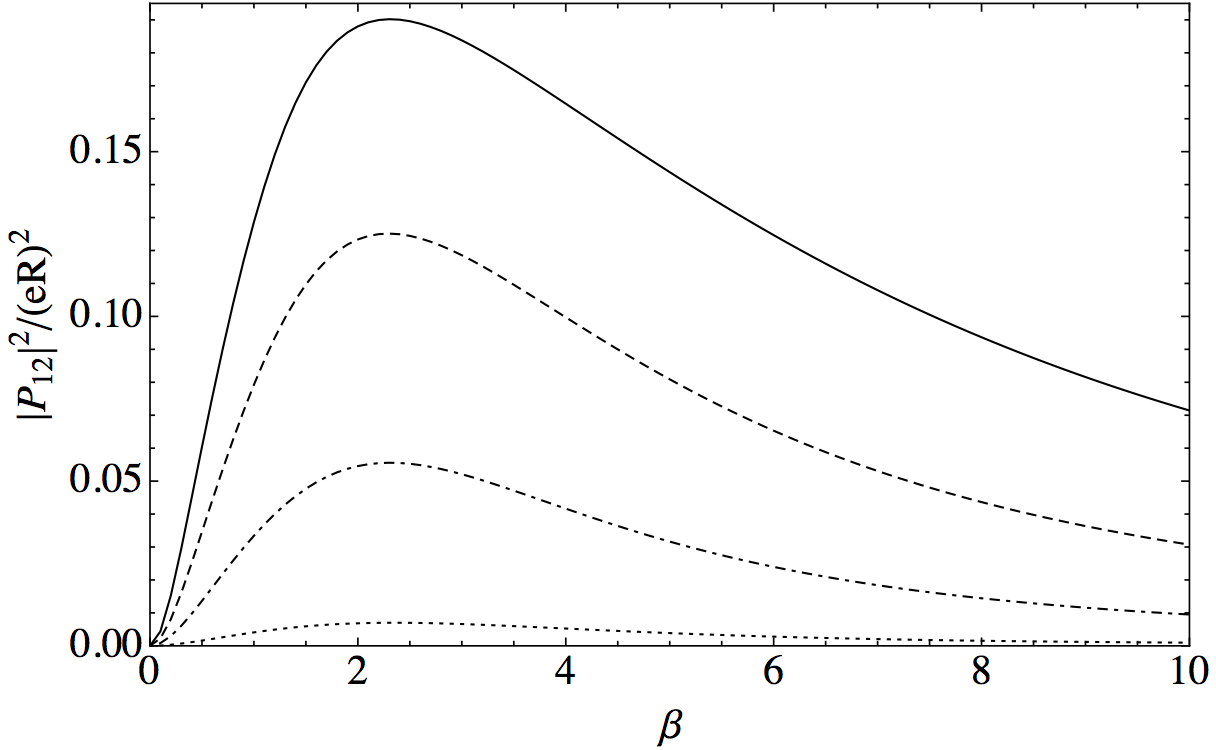}
\end{center}
\caption{\label{fig:dipolematrixelements12} Square of dimensionless transition dipole matrix element between first and second excited states, for the asymmetric system and $\theta=\pi/2$, plotted as a function of $\beta$. The solid line is $\gamma=1.007$, dashed line is $\gamma=1.005$, dot-dashed is $\gamma=1.003$, and dotted is $\gamma=1.001$.}
\end{figure}

Fig.~(\ref{fig:dipolematrixelements0102}) plots the dimensionless transition dipole matrix elements $|P_{01}|^2/(eR)^2$ at polarization angle $\theta=\pi/2$ and $|P_{02}|^2/(eR)^2$ at $\theta=0$, for different values of $\gamma$. The inset shows a schematic of the optical selection rules, with blue arrows representing allowed transitions for $\gamma=1$ and red dot-dashed arrows showing forbidden transitions, which become allowed when $\gamma \ne 1$. Asymmetry introduces the $\cos(\varphi)$ term in the potential \eqref{eq:potential} and removes the symmetry about $\varphi=\pi/2$, therefore coupling states with $\Delta m = \pm 1$. Formal solutions can then be written as
\begin{equation}
\Psi_{n}(\varphi) = \left \{
  \begin{aligned}
    & \frac{1}{\sqrt{\pi}}\sum_{j=0}A^{(n)}_{j}\cos(j\varphi), && n\ \text{even or}\ 0, \\
    &\frac{1}{\sqrt{\pi}}\sum\limits_{j=0}B^{(n)}_{j+1}\sin[(j+1)\varphi], && n\ \text{odd.}
  \end{aligned} \right.
  \label{eq:asymmetric states}
\end{equation} 
The matrix elements $P_{01}(\theta)$ and $P_{02}(\theta)$ maintain the same dependence on the polarization angle $\theta$, however decrease in magnitude with increasing asymmetry, as can be seen in Fig.~(\ref{fig:dipolematrixelements0102}). Optical transitions between the first and second states become allowed, with the corresponding matrix element given as
\begin{equation}
  P_{12}(\theta) = eR\sin(\theta)\sum_{j,k=0}\frac{A_j^{(2)}B_{k+1}^{(1)}}{2-\delta_{j,0}}\left(\delta_{j,k} - \delta_{j,k+2} \right).
    \label{eq:TransitionDipoleMatrixElementAsymmetric12}
\end{equation}
In the same way, all other optical transitions forbidden when $\gamma=1$ can be shown to become allowed when $\gamma \ne 1$, as is depicted schematically by the inset of Fig.~(\ref{fig:dipolematrixelements0102}). In Fig.~(\ref{fig:dipolematrixelements12}) we plot the matrix element $|P_{12}|^2/(eR)^2$ at $\theta=\pi/2$ as a function of $\beta$ for several values of $\gamma$. 

Switching the sign of the gate voltages ($\beta < 0$) for $\gamma=1$ is equivalent to the coordinate shift $V(\varphi) \rightarrow V(\varphi + \pi/2$) (up to an arbitrary constant), and the parities of the excited states about $\varphi=0$ and $\varphi=\pi/2$ are exchanged. Accordingly, while the matrix elements yield the same magnitudes, the polarization angle-dependence is rotated by $\pi/2$, i.e. $P_{01}(\theta)\propto \cos(\theta)$ and $P_{02}(\theta)\propto \sin(\theta)$. This dependence on $\theta$ is maintained for $\gamma \ne 1$; however, the highly tunable energy separation occurs between the first and second excited states, due to the resonant levels of the asymmetric DQW formed.

Controlling the polarization angle of incident radiation one can selectively excite transitions from the ground state to either of the next two excited states. Introducing asymmetry allows transitions between these two states. This opens the possibility for a double-gated QR to be viewed as a three-level system suitable for lasing between the first excited state and the ground state, driven by excitation from the ground state to the second excited state at a much higher frequency. This system offers control over the frequency of the lasing transition, which is typically in the THz range, via tuning the voltage on the gates.

\section{\label{sec:Conclusion}Conclusion}

In summary, we show that the system of a quantum ring between two gates is a tunable double quantum well. The energy separation between the ground state and first excited state for an electron in such a system can be tuned via the voltage on the gates. The problem is treated numerically via finite matrix diagonalization, which provides a good agreement with the perturbation theory and WKB method results in the limiting cases. Contrary to typical DQW heterostructures, dipole transitions in a double-gated QR induced by light polarized in the plane of the ring depend strongly on the polarization angle with respect to the gates. Consequently, we show that transitions between the ground state and the second excited state are allowed, with maximum rate of transitions for negatively (positively) charged gates occurring due to radiation polarized parallel (perpendicular) to the axis of the gates. Furthermore, the transition between the second and first excited states is forbidden in the particular case of a $\pi$-periodic potential, in contrast to planar DQW structures wherein this transition is always allowed. Due to the polarization angle-dependent selection rules, a double-gated QR may act as a three-level system capable of lasing between the highly tunable transition from first excited to ground state, with the differently polarized pump transition from the ground to second excited state occurring at higher frequency.

\begin{acknowledgments}
We acknowledge support from the Engineering and Physical Sciences Research Council (EPSRC) of the United Kingdom, via the EPSRC Centre for Doctoral Training in Metamaterials XM$^2$ (Grant No. EP/L015331/1). This work was also supported by the EU FP7 ITN NOTEDEV (Grant No. FP7-607521); EU H2020 RISE project CoExAN (Grant No. H2020-644076); FP7 IRSES projects CANTOR (Grant No. FP7-612285), QOCaN (Grant No FP7-316432), and InterNoM (Grant No FP7-612624). We are grateful to A. Shytov, E. Hendry and G. Nash for fruitful discussions.
\end{acknowledgments}

\end{document}